 \documentclass[twocolumn,amsmath,amssymb]{revtex4}
 \usepackage{graphicx} 
 
\begin{document} 
 
\title{Hadronic resonance gas and charged particle's $p_T$ spectra and elliptic flow in $\sqrt{s}$=200 GeV Au+Au collisions}
  
\date{\today}
  
\author{Victor Roy}
\email{victor@veccal.ernet.in}
\author{A. K. Chaudhuri} 
\email{akc@veccal.ernet.in} 
\affiliation{Variable Energy Cyclotron Centre, 1-AF, Bidhan Nagar,
Kolkata - 700 064, India} 
 \begin{abstract}
  
Charged particles $p_T$ spectra and elliptic flow in 0-60\% Au+Au collisions at RHIC are analyzed in a hydrodynamic model with hot, hadronic resonance gas in the initial state.  Physically conceivable hadronic resonance gas, thermalized in the time scale $\tau_i$=1 fm, at a (central) temperature  $T_i$=220 MeV,
with viscosity to entropy ratio $\eta/s$=0.24, reasonably well explains the $p_T$ spectra in all the collision centralities. Centrality dependence of elliptic flow however demands continual increase of viscosity to entropy ratio with centrality.
   \end{abstract}
 \maketitle  
 Recent experiments in Au+Au collisions at Relativistic Heavy Ion Collider (RHIC) \cite{BRAHMSwhitepaper},\cite{PHOBOSwhitepaper},\cite{PHENIXwhitepaper},\cite{STARwhitepaper}  produced convincing evidences that in central and mid central Au+Au collisions, a hot, dense, strongly interacting matter is created. Whether the matter can be identified as the lattice QCD \cite{Cheng:2007jq} predicted Quark Gluon Plasma(QGP) or not is still a question of debate. Confirmatory identification of the matter  with QGP requires complete elimination of the possibility that a hadronic  state is produced in the initial collisions. 
It is yet to be done. 
 Information about the initial state is always indirect. QGP is a transient state and even if produced, exists for a short time scale, it expands, cools and eventually transforms into hadrons. Hadrons are the experimental observables and any information about the initial state has to be obtained from the observed hadrons. Dynamical models are essential to extract information about the initial state from the experimental observables. 
  
Relativistic hydrodynamics is one of the few dynamical models which has been 
successfully applied in RHIC collisions to obtain information about the initial medium produced in Au+Au collisions. It is assumed that in a Au+Au collision a fireball is produced. Constituents of the fireball collide frequently to establish local thermal equilibrium sufficiently fast and after a certain time $\tau_i$, hydrodynamics become applicable. If the macroscopic properties of the fluid e.g. energy density, pressure, velocity etc. are known at the equilibration time $\tau_i$, the relativistic hydrodynamic equations can be solved to give the space-time evolution of the fireball till a given freeze-out condition such that interactions between the constituents become too weak to continue the evolution. 
Using suitable algorithm (e.g. Cooper-Frye) information at the freeze-out can be converted into particle spectra and can be directly compared with experimental data. Thus, hydrodynamics, in an indirect way, can characterize the initial state of the medium produced in heavy ion collisions. Hydrodynamic equations are closed only with an equation of state  and one can investigate the possibility of phase transition in the medium.

Relativistic hydrodynamics with QGP in the initial  state, thermalized in the time scale $\tau_i\approx $0.6 fm, at an initial central energy density $\varepsilon_0\approx$ 30 $GeV/fm^3$, could explain 
a host of experimental data produced in $\sqrt{s}$=200 GeV Au+Au collisions  \cite{QGP3}. In 
an alternative to the initial  QGP state, namely the hot hadronic resonance gas (HRG),  description to the experimental data gets much poorer \cite{Kolb:2000fha}. Also, to reproduce the experimental multiplicity,
hot hadronic resonance gas is required to be initialized at very high temperature, e.g. $T_i\approx$ 270 MeV. Density of HRG grows  rapidly with temperature. At $T\approx$ 270 MeV, density of hadrons is large, $\rho_{had} \sim$ 4 $fm^{-3}$. At  such a large density hadrons will overlap extensively and it is difficult to believe that they could retain their individual identity. It does appear that a consistent description of RHIC data could not be obtained in a hadronic resonance gas model.
However, all the analysis of RHIC data with hadronic resonance gas  in the initial state was performed in the ideal hydrodynamic limit. 
In recent years, there is much progress in practical application of viscous hydrodynamics \cite{Teaney:2003kp}, \cite{Muronga:2003ta}, \cite{Heinz:2005bw},\cite{Chaudhuri:2006jd}, \cite{Song:2007fn}, \cite{Dusling:2007gi},\cite{Chaudhuri:2008ed}, \cite{Chaudhuri:2008sj}, \cite{Chaudhuri:2008je}, \cite{Song:2008si}, \cite{Romatschke:2007mq}, \cite{Chaudhuri:2009uk}, \cite{Chaudhuri:2009hj}, \cite{Chaudhuri:2009ud}, \cite{Roy:2010qg}. Several codes were developed to numerically solve causal dissipative hydrodynamics and have been applied    successfully to analyze experimental data at RHIC collisions with QGP in the initial state. 
However, we have not come across any analysis of RHIC data  with 'viscous' hadronic resonance gas in the initial state. 
Viscous effects can be large in hadronic resonance gas. 
Model calculations \cite{Gorenstein:2007mw},\cite{Demir:2008tr},\cite{Muronga:2003tb},\cite{Pal:2010es}   indicate that viscosity to entropy ratio of a hadronic resonance gas can be considerably larger than the ADS/CFT limit, $\eta/s \geq 1/4\pi$  \cite{Policastro:2001yc}. 
Since entropy is generated during evolution, unlike an 'ideal' HRG, a 'viscous' HRG can be initialized at a lower temperature such that hadrons retain their identity yet reproduce the experimental multiplicity.
Indeed, in one of the earliest applications of one dimensional viscous hydrodynamics, it was shown that WA80 single photon spectra,  though over predicted in an ideal hadron gas evolution,  is reasonably well explained when viscous effects are accounted for \cite{Chaudhuri:1995se}.   
 To exclude the possibility of  hadron gas formation  in initial Au+Au collisions, instead of QGP, it is important that RHIC data are analyzed in a viscous hadronic resonance gas model. 

We assume that in Au+Au collisions a baryon free, hot hadronic resonance gas, comprising all the hadron resonances with mass $m_{res} \leq$2.5 GeV is produced.
The hot, hadronic resonance gas is assumed to thermalize in the time scale $\tau_i$=1 fm. We also neglect all the dissipative effects except for the shear viscosity. The space time evolution of the fluid is obtained by solving, 
  
\begin{eqnarray}  
\partial_\mu T^{\mu\nu} & = & 0,  \label{eq1} \\
D\pi^{\mu\nu} & = & -\frac{1}{\tau_\pi} (\pi^{\mu\nu}-2\eta \nabla^{<\mu} u^{\nu>}) \nonumber \\
&-&[u^\mu\pi^{\nu\lambda}+u^\nu\pi^{\nu\lambda}]Du_\lambda. \label{eq2}
\end{eqnarray}

Eq.\ref{eq1} is the conservation equation for the energy-momentum tensor, $T^{\mu\nu}=(\varepsilon+p)u^\mu u^\nu - pg^{\mu\nu}+\pi^{\mu\nu}$, 
$\varepsilon$, $p$ and $u$ being the energy density, pressure and fluid velocity respectively. $\pi^{\mu\nu}$ is the shear stress tensor (we have neglected bulk viscosity and heat conduction). Eq.\ref{eq2} is the relaxation equation for the shear stress tensor $\pi^{\mu\nu}$.   
In Eq.\ref{eq2}, $D=u^\mu \partial_\mu$ is the convective time derivative, $\nabla^{<\mu} u^{\nu>}= \frac{1}{2}(\nabla^\mu u^\nu + \nabla^\nu u^\mu)-\frac{1}{3}  
(\partial . u) (g^{\mu\nu}-u^\mu u^\nu)$ is a symmetric traceless tensor. $\eta$ is the shear viscosity and $\tau_\pi$ is the relaxation time.  It may be mentioned that in a conformally symmetric fluid relaxation equation can contain additional terms  \cite{Song:2008si}. Assuming longitudinal boost-invariance, 
the equations are solved  with the code 'AZHYDRO-KOLKATA'\cite{Chaudhuri:2008je} in ($\tau=\sqrt{t^{2}-z^{2}},x,y,\eta=\frac{1}{2}ln\frac{t+z}{t-z}$) coordinates.  
  
Solution of Eq.\ref{eq1} and \ref{eq2} require initial energy density, velocity
distribution in the transverse plane at the initial time. We assume that at the initial time $\tau_i$=1 fm,  initial fluid velocity is zero, $v_x(x,y)=v_y(x,y)$=0.   The initial energy density in  an impact parameter ${\bf b}$ collision is assumed to be  distributed as \cite{QGP3},

\begin{equation} \label{eq3}
\varepsilon({\bf b},x,y) = \varepsilon_0[(1-x)N_{part} ({\bf b},x,y)+ x N_{coll}({\bf b},x,y)]
\end{equation}  
   
\noindent where $N_{part}({\bf b},x,y)$ and $N_{coll}({\bf b},x,y)$ are the transverse profile of participant numbers and binary collision numbers respectively. $N_{part}({\bf b},x,y)$ and $N_{coll}({\bf b},x,y)$ can be calculated in a Glauber model. $x$ 
in Eq.\ref{eq3} is the fraction of hard scattering. Most of the hydrodynamic simulations are performed with hard scattering fraction $x$=0.25 or  0.13 \cite{QGP3},\cite{Hirano:2009ah}.
Recently, in  \cite{Roy:2010zd}, it was shown that in collisions beyond 0-10\% centrality, simultaneous description of charged particles $p_T$ spectra and elliptic flow are best obtained with hard scattering fraction $x=0$.   Only in 0-10\% centrality collisions, hard scattering fraction $x=1$ is preferred. In the following, in all the collision centralities, we assume hard scattering fraction $x$=0, understanding that we may under predict elliptic flow in very central collisions.
$\varepsilon_0$ in Eq.\ref{eq3} is the central energy density of the fluid in an impact parameter ${\bf b}$=0 collision. Generally, $\varepsilon_0$ is obtained by fitting  experimental data, e.g. multiplicity, $p_T$ spectra etc. 
We however fix $\varepsilon_0$  to the highest possible value for a 'physically' conceivable hadronic resonance gas.
For hadron size $\approx$ 0.5 fm, limiting hadron density (such that hadrons are not overlapped) is $\rho^{had}_{limit}=1/V \approx$ 2 $fm^{-3}$. $\rho^{had}_{limit}$=2 $fm^{-3}$ corresponds to limiting temperature or energy density, $T_{limit}$= 220 MeV, $\varepsilon_{limit}$=5.1 $GeV/fm^3$. We fix the central energy density at this limiting value,  $\varepsilon_0=\varepsilon_{limit}$=5.1 $GeV/fm^3$. 
Recent lattice simulations  \cite{Cheng:2007jq}, with almost physical quark masses ($m_\pi\approx$220 MeV), predicts a confinement-deconfinement transition  temperature $T_c=196 \pm 3$ MeV. Hadrons can exists at $T\approx$200 MeV. Existence of HRG at $\sim$10\% higher temperature  $T$=220MeV is a definite possibility.

  Dissipative hydrodynamics also require initialization of the shear stress tensor $\pi^{\mu\nu}$ as well as the relaxation time $\tau_\pi$. We assume that shear stress tensors are initialized at the boost-invariant values, $\pi^{xx}=\pi^{yy}=2\eta(x,y)/3\tau_i$, $\pi^{xy}=0$ \cite{Chaudhuri:2008je}.  For the relaxation time $\tau_\pi$, we use the
Boltzmann estimate, $\tau_\pi=6\eta/4p \approx \frac{6}{T}\frac{\eta}{s}$. Finally, hydrodynamic models   require a freeze-out condition. We assume that the fluid freeze-out at a fixed temperature $T_F$=110 MeV.

\begin{figure}[t]
\center
 \resizebox{0.45\textwidth}{!}{%
  \includegraphics{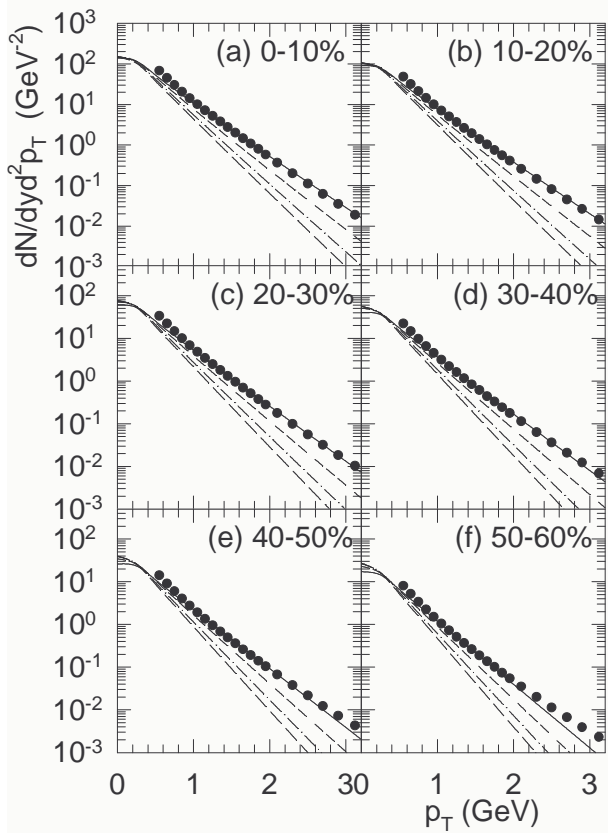}
}
\caption{In six panels, PHENIX measurements \cite{Adler:2003au} for charged particles $p_T$ spectra in 0-10\%, 10-20\%, 20-30\%, 30-40\%, 40-50\% and 50-60\% Au+Au collisions are shown. The dashed, dash-dotted,  short dashed and solid lines are simulated spectra from evolution hot hadronic resonance gas with $\eta/s$=0, $\eta/s$=0.08 and $\eta/s$=0.16 and $\eta/s$=0.24 respectively.} .\label{F1}
\end{figure}

\begin{figure}[t]
\center
 \resizebox{0.45\textwidth}{!}{%
  \includegraphics{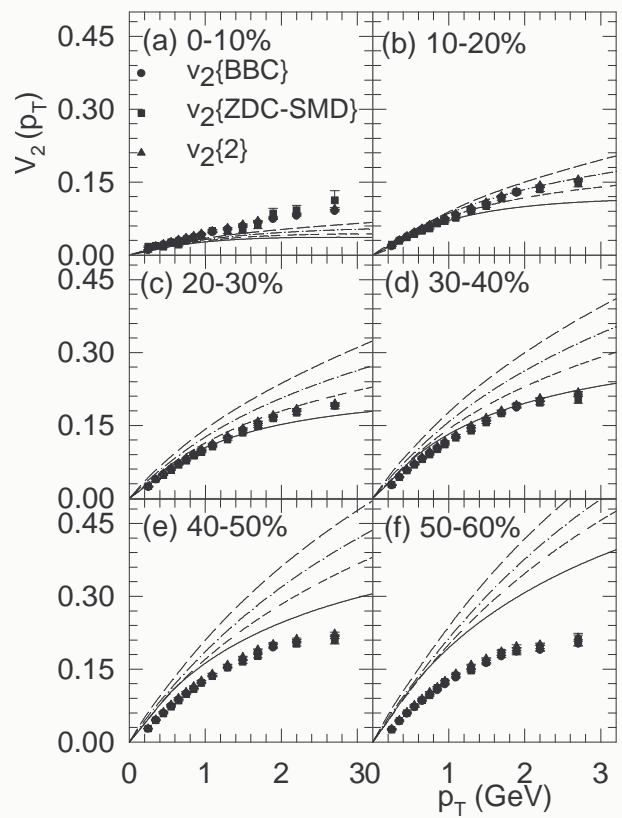}
}
\caption{In six panels, PHENIX measurements \cite{Afanasiev:2009wq}  for charged particles elliptic flow in 0-10\%, 10-20\%, 20-30\%, 30-40\%, 40-50\% and 50-60\% Au+Au collisions are shown. The dashed, dash-dotted,  short dashed and solid lines are simulated flow from evolution hot hadronic resonance gas with $\eta/s$=0, $\eta/s$=0.08 and $\eta/s$=0.16 and $\eta/s$=0.24 respectively.}
 \label{F2}
\end{figure}

Only unspecified parameter in the model is the viscosity coefficient $\eta$. We assume that throughout the evolution,
the ratio of viscosity to entropy density $\eta/s$ remain a constant and
 simulate Au+Au collisions for four values of $\eta/s$, (i) $\eta/s$=0 (ideal fluid), (ii) $\eta/s=1/4\pi$=0.08 (ADS/CFT lower limit of viscosity) and (iii) $\eta/s$=0.16 and (iv) $\eta/s$=0.24. Assumption of constant $\eta/s$ fixes the variation of viscosity with temperature, $\eta \propto s \propto  T^3$. In Fig.\ref{F1} and \ref{F2},
simulated   charged particles $p_T$ spectra and elliptic flow are compared with experimental  data. In Fig.\ref{F1}, in six panels, PHENIX data \cite{Adler:2003au} for charged particles $p_T$ spectra in 0-10\%, 10-20\%, 20-30\%, 30-40\%, 40-50\% and 50-60\% Au+Au collisions are shown.
The dashed, dashed-dot, short dashed and solid lines in the figure are simulated spectra from evolution of hot hadronic resonance gas, with viscosity to entropy ratio $\eta/s$=0, 0.08, 0.16 and 0.24 respectively. If viscous effects are neglected, hot hadronic resonance gas, initialized with central temperature $T_i$=220 MeV, do not explains the data, the data are largely under predicted. For example, at $p_T\approx$1.75 GeV, in all the collision centralities, ideal hot hadronic gas under predict  PHENIX data by a factor of $\sim$6. Data at higher $p_T$ are even more under predicted. When viscous effects are included, 
particle yield increases, more at high $p_T$ than at low $p_T$ and discrepancy with experiment and simulated spectra diminishes. At $p_T\approx$1.75 GeV, data are under predicted,    by a factor of $\sim$ 4 in evolution of minimally viscous HRG ($\eta/s$=0.08) and by a factor of $\sim$ 2  in evolution of HRG with $\eta/s$=0.16. Experimental data are reasonably reproduced with viscosity to entropy ratio $\eta/s$=0.24. 
Indeed, in all the centrality ranges of collisions, simulated spectra from evolution of hadronic resonance gas with viscosity to entropy ratio $\eta/s$=0.24,   agree with experimental within $\sim$10\%.   It is very interesting to note that the $\eta/s$=0.24, obtained from the analysis     is in close agreement with theoretical estimates of viscosity to entropy ratio of a hot hadronic resonance gas, $\eta/s$=0.24-0.30   \cite{Gorenstein:2007mw,Pal:2010es}.

The analysis clearly indicate that  
at least for the charged particles $p_T$ spectra , it is not necessary that QGP fluid is produced in  $\sqrt{s}$=200 GeV Au+Au collisions. 'Physically conceivable' hot hadronic resonance gas, with viscosity to entropy ratio $\eta/s$=0.24, could reproduce the spectra. 
Elliptic flow analysis on the other hand gives a different result. 
In fig.\ref{F2}, PHENIX measurements \cite{Afanasiev:2009wq}  for charged particles elliptic flow in 0-10\%, 10-20\%, 20-30\%, 30-40\%, 40-50\% and 50-60\% Au+Au collisions are shown. PHENIX collaboration measured charged particles $v_2$ upto $p_T\approx$ 8 GeV. In Fig.\ref{F2},   measurements upto $p_T$=3 GeV are shown only. Hydrodynamic models are not well suited for high $p_T$ particles.
 In order to study non-flow
effects that are not correlated with the reaction plane, as well as fluctuations of $v_2$, PHENIX collaboration obtained $v_2$ from two independent analysis,
(i) event plane method from two independent subdetectors, $v_2\{BBC\}$ and 
$v_2 \{ZDC-SMD\}$ and (ii) two particle cumulant $v_2\{2\}$.  $v_2\{2\}$ from two particle cumulant and $v_2\{BBC\}$ or $v_2\{ZDC-BBC\}$ from event plane methods agree within the systematic error.   It may also be mentioned here that   $v_2\{2\}$ in PHENIX is lower than that $v_2\{2\}$in STAR measurements,  but they agree within the systematic error. All the three measurements of $v_2$ are shown in Fig.\ref{F2}. As before, the dashed, dashed-dotted, short-dashed and solid lines in Fig.\ref{F2} are simulated flow with $\eta/s$=0, 0.08, 0.16 and 0.24 respectively. Unlike the $p_T$ spectra, centrality dependence of elliptic flow are not explained with a single value for viscosity to entropy ratio. Data demand more viscous fluid in more peripheral collisions. For example, in 0-10\% centrality collisions, experimental flow are under predicted
  in ideal HRG. With viscous HRG, flow is even more under predicted. We have assumed participant scaling for the initial energy density. As mentioned earlier, 
very central collisions prefer binary collision number scaling of initial energy density \cite{Roy:2010zd}. Spatial  eccentricity is comparatively large in binary collision scaling  and elliptic flow in 0-10\% collision will be better explained if initial energy density scales with binary collision numbers.
In 10-20\% centrality collisions, experimental flow is under predicted with $\eta/s$=0.16 and 0.24. Data demand HRG with  viscosity in the range $\eta/s$=0-0.08. Elliptic flow in 20-30\% centrality collisions prefers more viscous HRG, $\eta/s\approx$0.16. Hadronic resonance gas with $\eta/s\approx$ 0.24 approximately explains flow in 30-40\% centrality collisions.   Elliptic flow in 40-50\% and 50-60\%
Au+Au collisions demand HRG with $\eta/s > 0.24$. Continual increase of viscosity with centrality can be understood. In a hadronic resonance gas viscosity to entropy ratio increases with decreasing temperature. Initial temperature of the fluid also decreases as the collisions become more and more peripheral. For example, central temperature of the fluid is $T_i\approx$ 219 MeV in a 0-10\% collision. In a   30-40\% collision, central temperature is $T_i\approx$210 MeV. Hadronic fluid will be more viscous in a peripheral collision than in a central collision. Elliptic flow, being a sensitive observable can detect the small variation in temperature. 

It is however difficult to explain why the stated change in temperature  is not reflected in the $p_T$ spectra. Though centrality dependence of elliptic flow require continual increase of viscosity,   charged particles $p_T$ spectra definitely do not demand such increase .
Rather the $p_T$ spectra are well explained with a fixed value for viscosity, $\eta/s\approx$0.24(see Fig.\ref{F1}). It appear that a viscous hydrodynamic model, with hadronic resonance gas in the initial state, is incapable of simultaneous explanation of centrality dependence of charged particles $p_T$ spectra and elliptic flow.  
We may note here that 
hydrodynamics, with QGP in the initial state, also require continual increase of viscosity to explain the centrality dependence of elliptic flow \cite{Chaudhuri:2009hj}, particle spectra are however are explained with a fixed viscosity to entropy \cite{Roy:2010qg}. 

Before we summarise, we note that we have used some specific initial conditions for the hydrodynamical model, e.g. initial (central) energy density $\varepsilon_i$= 5.1 $GeV/fm^3$,  initial or thermalization time $\tau_i$=1 fm, freeze-out temperature $T_F$=110 MeV. The initial energy density was fixed from physical requirement  that constituents of the hadronic resonance gas do not overlap extensively and loss their identity. Hydrodynamic model analysis of RHIC data with QGP in the initial state indicate thermalization time for QGP $\tau_i$=0.6 fm. Compared to QGP, a hadronic resonance gas is expected to take longer time to achieve thermalization and the thermalization time for the hadronic resonance gas was fixed to the canonical value $\tau_i$=1 fm. Similarly, freeze-out temperature $T_F$=110 MeV was indicated in the hydrodynamic model analysis of RHIC data \cite{QGP3}.  All physically supported parameters set were not examined.

In summary, we have studied the possibility of explaining charged particle's $p_T$ spectra and elliptic flow without invoking Quark-Gluon-Plasma. 
We assume that  
a hot hadronic resonance gas is formed in initial Au+Au collisions. The hot hadronic gas thermalizes in the time scale $\tau_i\approx$ 1 fm to central temperature
$T_i\approx$ 220 MeV. While an ideal hadronic resonance gas do not explain the charged particle $p_T$ spectra (data are largely under predicted),  a viscous hadronic gas, with viscosity to entropy ratio $\eta/s\approx$0.24, explains the $p_T$ spectra in all the   centrality
ranges of collisions. Elliptic flow data however are more
sensitive and demand more viscous HRG in peripheral collisions than in central collisions. We conclude that with the initialization and freeze-out parametrization presented here, a viscous hydrodynamic model containing only hadronic resonance gas, is incapable of simultaneous explanation of centrality dependence of charged particles $p_T$ spectra and elliptic flow in $\sqrt{s}$=200 GeV Au+Au collisions at RHIC.

\end{document}